%% file: paper.tex
\documentclass[submission,copyright,creativecommons]{eptcs}
\usepackage[dvipsnames]{xcolor}
\providecommand{\event}{LFMTP 2019} 
\usepackage{breakurl}             
\usepackage{underscore}           

\usepackage{scalerel}
\usepackage{tikz}
\usetikzlibrary{svg.path}

\definecolor{orcidlogocol}{HTML}{A6CE39}
\tikzset{
  orcidlogo/.pic={
    \fill[orcidlogocol] svg{M256,128c0,70.7-57.3,128-128,128C57.3,256,0,198.7,0,128C0,57.3,57.3,0,128,0C198.7,0,256,57.3,256,128z};
    \fill[white] svg{M86.3,186.2H70.9V79.1h15.4v48.4V186.2z}
                 svg{M108.9,79.1h41.6c39.6,0,57,28.3,57,53.6c0,27.5-21.5,53.6-56.8,53.6h-41.8V79.1z M124.3,172.4h24.5c34.9,0,42.9-26.5,42.9-39.7c0-21.5-13.7-39.7-43.7-39.7h-23.7V172.4z}
                 svg{M88.7,56.8c0,5.5-4.5,10.1-10.1,10.1c-5.6,0-10.1-4.6-10.1-10.1c0-5.6,4.5-10.1,10.1-10.1C84.2,46.7,88.7,51.3,88.7,56.8z};
  }
}

\newcommand\orcidicon[1]{\href{https://orcid.org/#1}{\mbox{\scalerel*{
\begin{tikzpicture}[yscale=-1,transform shape]
\pic{orcidlogo};
\end{tikzpicture}
}{|}}}}

\usepackage{basics,ded}

\usepackage[utf8]{inputenc}
\usepackage{amsmath,amssymb,relsize}
\usepackage{xspace}
\usepackage{mytikz}

\input{listings}
\hypersetup{linkcolor=red,citecolor=blue,urlcolor=gray,colorlinks,breaklinks,bookmarksopen,bookmarksnumbered}

\usepackage[hide]{ed}
\newif\ifshort\shorttrue

\title{Rapid Prototyping Formal Systems in MMT: 5 Case Studies}

\author{Dennis Müller \institute{Computer Science, FAU Erlangen-N\"urnberg, Germany}\email{d.mueller@kwarc.info} \orcidicon{0000-0002-4482-4912} \and
Florian Rabe \institute{Computer Science, FAU Erlangen-N\"urnberg, Germany} \institute{LRI Paris, France} \email{florian.rabe@fau.de}\orcidicon{0000-0003-3040-3655}
}

\def\authorrunning{D. Müller and F. Rabe}
\def\titlerunning{Rapid Prototyping Formal Systems in MMT}

\input{macros}

\begin{document}

\maketitle

\begin{abstract}
Logical frameworks are meta-formalisms in which the syntax and semantics of object logics and related formal systems can be defined.
This allows object logics to inherit implementations from the framework including, e.g., parser, type checker, or module system.
But if the desired object logic falls outside the comfort zone of the logical framework, these definitions may become cumbersome or infeasible.

Therefore, the MMT system abstracts even further than previous frameworks: it assumes no type system or logic at all and allows its kernel algorithms to be customized by almost arbitrary sets of rules.
In particular, this allows implementing standard logical frameworks like LF in MMT.
But it does so without chaining users to one particular meta-formalism: users can flexibly adapt MMT whenever the object logic demands it.

In this paper, we present a series of case studies that do just that, defining increasingly complex object logics in MMT.
We use elegant declarative logic definitions wherever possible, but inject entirely new rules into the kernel when necessary.
Our experience shows that the MMT approach allows deriving prototype implementations of very diverse formal systems very easily and quickly.
\end{abstract}

\section{Introduction and Related Work}
 \input{intro}

\section{Preliminaries}\label{sec:prelim}
\input{syntax}

\lstMakeShortInline[language=scala]*

\section{Dependently-Typed Higher-Order Logic}\label{sec:dhol}
\input{dhol}

\section{LF Modulo Rewriting}\label{sec:rewrite}
\input{rewriting}

\section{External Side Conditions}\label{sec:lock}
\input{locks}

\section{Linear Logic}\label{sec:linear}
\input{linear}

\section{Homotopy Type Theory}\label{sec:hott}
\input{hott}

\lstDeleteShortInline*

\section{Conclusion and Future Work}\label{sec:conc}
\input{conc}

\bibliographystyle{eptcs}

\bibliography{biblio}

\end{document}

%% file: listings.tex
\usepackage{listings}

\lstnewenvironment{code}[1][mathescape=false]{\lstset{language=XML, #1}}{}

\newcommand{\delima}{\mid\negmedspace \mid}
\newcommand{\delimb}{\delima\negmedspace \delima}
\newcommand{\GS}{\ensuremath{\textcolor{RedViolet}{\delimb\negmedspace\delimb}}}
\newcommand{\RS}{\ensuremath{\textcolor{Green}{\delimb}}}
\newcommand{\US}{\ensuremath{\textcolor{Turquoise}{\delima}}}

\lstdefinestyle{mmt}{	
	keywordstyle=\color{Emerald}
}
\lstdefinelanguage{mmt}{
	morekeywords={theory,view,include,namespace,rule,role,implicit,import,structure},
	morecomment=[l][\color{Gray}]{//},
}
\lstnewenvironment{mmtcode}[1][language=mmt]{\lstset{language=mmt,style=mmt,inputencoding=utf8,escapechar=\&,#1}}{}
\lstnewenvironment{nmmtcode}[1][language=mmt]{\lstset{xleftmargin=1.2em,numbers=left,language=mmt,style=mmt,inputencoding=utf8,escapechar=\&,#1}}{}
\lstdefinestyle{scala}{	
	keywordstyle=\color{BurntOrange}
}
\lstdefinelanguage{scala}{
	emphstyle=\underbar,
	classoffset=0,keywordstyle=\bfseries\color{BurntOrange},
	morekeywords={case,class,object,trait,extends,match,if,else,def,implicit,import,val,var,this,override,with,return,abstract},
	classoffset=1,keywordstyle=\color{blue},
	morekeywords={OMA,OMS,OMBIND,OMV,OMID,Context,VarDecl,Term,LocalName,GlobalName,MPath,Solver,Stack,Rule,
	History,Inhabitable,Subtyping,Equality,Typing,Common,Judgment,FormationRule,CheckingRule,IntroductionRule,
	EliminationRule,InferenceRule,TypingRule,ComputationRule,TypeBasedEqualityRule,TermBasedEqualityRule,
	TermHeadBasedEqualityRule,Continue,CheckingCallback,Simplifiability,Simplify,RecurseOnly,SubtypingRule,
	VarianceRule,TypeBasedSolutionRule,DerivedDeclaration,StructuralFeature,Elaboration,FormatBasedExtension,
	Marker,ExtendedCheckingEnvironment,Module,Universe,InhabitableRule,UniverseRule,ElementContainer,Declaration,
	OMMOD,ContentPath,RealizedType,SemanticType,StandardNat,Constant,Theory,StructuralElement,ChangeListener,
	RuleConstant,Extension},
	classoffset=2,keywordstyle=\color{PineGreen},
	morekeywords={Option,Some,None,Boolean,String,List,Int,true,false,Nil,BigInt,Double},
	classoffset=0,
	morecomment=[l][\color{Gray}]{//},
	morecomment=[s][\color{Gray}]{/*}{*/},
	morecomment=[s][\color{OliveGreen}]{/**}{**/},
	morestring=[b][\color{Purple}]",
}
\lstnewenvironment{scalacode}[1][language=scala]{\lstset{language=scala,style=scala,showstringspaces=false,inputencoding=utf8,#1}}{}
\lstnewenvironment{nscalacode}[1][language=scala]{\lstset{xleftmargin=1.2em,numbers=left,language=scala,style=scala,showstringspaces=false,inputencoding=utf8,#1}}{}

%% file: macros.tex
\newcommand{\vmk}[1]{V#1}
\newcommand{\amk}[1]{#1}
\newcommand{\cons}[3]{#1\left(#2;\;#3\right)}

\newcommand{\key}[1]{\mathtt{#1}}
\newcommand{\rldecl}[1]{\key{rule}\,#1}

\newcommand{\llfp}{LLF$_\mathcal{P}$\xspace}

\newcommand\mylink[3]{#3}

\newcommand{\sub}[2]{\mylink{Bittersweet}{subst}{\left[ \vphantom{#1 /_{#2}} \right. } #1 \mylink{Bittersweet}{subst}{/}_{#2}\mylink{Bittersweet}{subst}{\left. \vphantom{#1 /_{#2}} \right]}}
\renewcommand{\bnf}[1]{{\color{red}#1}}

\newcommand{\cn}[1]{\ensuremath{\mathtt{#1}}}

\makeatletter
	\newenvironment{scriptaligned}[1][c]
	{\, \hbox\bgroup
		\fontsize{ \sf@size}{\dimexpr \sf@size pt+1pt}\selectfont
		$\! \aligned[#1]
	}{ \endaligned$\egroup}
\makeatother

\newcommand{\trule}[2] {\dfrac{\begin{scriptaligned}#1\end{scriptaligned}}{\begin{scriptaligned}#2\end{scriptaligned}}}

\newcommand{\judgmentscolor}{RedViolet}

\newcommand{\judg}[2]{#1 \mylink{\judgmentscolor}{judg}{\vdash} #2}
\newcommand{\gjudg}[1]{\judg\Gamma{#1}}

\newcommand{\infertype}[2]{ #1 \mylink{\judgmentscolor}{judg}{\Rightarrow} #2 }
\newcommand{\hastype}[2]{ #1 \mylink{\judgmentscolor}{judg}{\Leftarrow} #2 }
\newcommand{\judgequt}[2]{ #1 \mylink{\judgmentscolor}{judg}{\equiv} #2}
\newcommand{\judgeq}[3]{\judgequt{#1}{#2}\;\mylink{\judgmentscolor}{judg}{:}#3}
\newcommand{\judgeqc}[2]{ #1 \mylink{\judgmentscolor}{judg}{\rewrites} #2}

\newcommand{\judginh}[1]{#1\;\mylink{\judgmentscolor}{judg}{\mathtt{inh}}}

\newcommand{\lfcolor}{ForestGreen}

\newcommand{\type}{\mylink{\lfcolor}{lfpi}{\ensuremath{\mathtt{type}}}\xspace}
\newcommand{\kind}{\mylink{\lfcolor}{lfpi}{\ensuremath{\mathtt{kind}}}\xspace}

\DeclareMathOperator*\lfpisymm{\mylink{\lfcolor}{lfpi}{\ensuremath{\prod}}}
\newcommand\lfpi[2]{\lfpisymm_{#1}#2}

\newcommand\lfarrowsym{\mylink{\lfcolor}{lfpi}{\ensuremath{\to}}}
\newcommand\lfarrow[2]{#1\; \lfarrowsym\; #2}

\newcommand{\wtypecolor}{Melon}

\DeclareMathOperator*{\wtypesymm}{\mylink{\wtypecolor}{wtype}{\textsf{\larger W}}}
\newcommand{\wtypesym}{\ensuremath{\wtypesymm}}
\newcommand{\wtype}[2]{\wtypesym_{#1} #2}
\newcommand{\wsupsym}{\mylink{\wtypecolor}{wsup}{\ensuremath{\sup}}\xspace}

\newcommand{\wsupu}[4]{
\wsupsym_{#1}\mylink{\wtypecolor}{wsup}{\left\{\vphantom{#2 #3 #4}\right. } #2 \mylink{\wtypecolor}{wsup}{\Longrightarrow}_{#4} #3\mylink{\wtypecolor}{wsup}{\left.\vphantom{#2 #3 #4}\right\}} }

\newcommand{\wrecsym}{\ensuremath{\mylink{\wtypecolor}{wrec}{\textsf{rec}}}\xspace}
\newcommand{\wrec}[7]{\wrecsym\mylink{\wtypecolor}{wrec}{ \left( \vphantom{#2} \right.}#2\mylink{\wtypecolor}{wrec}{\left. \vphantom{#2}\right)}\mylink{\wtypecolor}{wrec}{\left\{ \left( \vphantom{#4 #5 #6} \right. \vphantom{#7} \right. } #4,#5,#6\mylink{\wtypecolor}{wrec}{\left. \vphantom{#4 #5 #6} \right)} \mylink{\wtypecolor}{wrec}{\Longrightarrow}_{#3} #7 \mylink{\wtypecolor}{wrec}{\left. \vphantom{#4 #5 #6 #7} \right\} } }

\usepackage{mdframed,realboxes}
\newenvironment{framed}{\begin{mdframed}\begin{center}}{\end{center}\end{mdframed}}

%% file: intro.tex
\subparagraph*{Motivation}
Despite its potential and successes, the automation of formal systems proceeds very slowly because designing them, implementing them, and scaling these implementations to practical tools is extremely difficult and time-consuming.
\textbf{Logical frameworks} provide meta-logics in which the syntax and semantics of object logics can be defined.
They are invaluable for \textbf{experimentation}: they allow rapidly prototyping formal systems, often to the extent of reducing the design-implement-scale process from person-years to person-days, and thus can speed up the feedback loop by orders of magnitude.

Features realized generically in logical frameworks include reasoning about object logics (Twelf \cite{twelf}, Abella \cite{abella}), interactive theorem proving (Isabelle \cite{isabelle}), concurrency (CLF \cite{clf}), reasoning about contexts (Beluga \cite{beluga}), rewriting (Dedukti \cite{dedukti}), side conditions (\llfp \cite{llfp}), or integration with proof assistants (Hybrid \cite{hybrid}).
Moreover, many logic-specific systems are investigating how to allow users to experiment with system behavior inside the system, e.g., via meta-programming (for Idris in \cite{idrisreflection}, for Lean in \cite{leanreflection}) or unification hints (for Coq in \cite{coqssreflect}, for Matita in \cite{matitahints}).


Within this group, \mmt \cite{rabe:recon:17} takes an extreme approach: it tries to systematically avoid any commitment to logical foundations while still allowing useful generic results \cite{RK:mmt:10,rabe:howto:14,rabe:mmtabs:13}.
All kernel algorithms of \mmt are parametrized by a set of rules, which are programmed in the underlying programming language (Scala).
The set of rules to use is collected from the current context so that entirely different kernel behaviors can coexist in the same development.
This makes \mmt so flexible that other logical frameworks can be conveniently developed inside it.
(This gave rise to the name \mmt as an abbreviation of meta-meta-theory/tool.)

This design makes \mmt most similar to the ELPI framework \cite{elpi}.
Both use an untyped expression language, into which object language syntax is embedded, and a programming language for writing rules.
But while ELPI uses the same language ($\lambda$-Prolog) for defining syntax and rules, \mmt uses a purely declarative language for the syntax and a general-purpose programming language for the rules.
Thus, it offers more freedom for rule definitions than ELPI at the price of disconnecting syntax and rules.
This freedom allows users fine-grained control over, e.g., error-reporting or side conditions, and enables the use of support tools like debuggers and IDEs, but makes it harder to reason about the rules.

%

\subparagraph*{Overview and Contribution}
Our objective is to evaluate the usability of \mmt by conducting several rapid prototyping experiments in it.
After sketching the basics of \mmt language in Sect.~\ref{sec:prelim}, using the dependent type theory LF \cite{lf} as a ``hello-world example'', the subsequent sections describe one case study each that extends LF with increasingly complex features: dependently-typed higher-order logic (DHOL) in Sect.~\ref{sec:dhol}, Dedukti-style rewriting in Sect.~\ref{sec:rewrite}, \llfp types as described in \cite{llfp} in Sect.~\ref{sec:lock}, linear logic as defined by the resource semantics in Sect.~\ref{sec:linear}, and homotopy type theory (HoTT) in Sect.~\ref{sec:hott}.

Our main contribution is to answer positively the question whether it is possible to build a framework that allows prototyping such a variety of features both elegantly and rapidly.

Regarding elegance: all prototypes involve only little non-declarative parts (i.e., rules programmed in Scala), and that code is orders of magnitude shorter and arguably easier to understand than a from-scratch implementation would be.

Regarding rapidness, we have tried to estimate the time effort for each case study.
DHOL took a few person-hours, rewriting about one person-week (distributed over several years), \llfp took two evenings, linear logic took two days (building on an existing failing case study in pure LF).
These were done by Rabe, who also built \mmt.
HoTT, combining various typing features, was done by M\"uller as a PhD student and in its first development -- with little to no prior experience with either MMT or type theories --- took several weeks over the course of roughly a year. A recent complete reimplementation was done in a few afternoons.

Each case study is published here for the first time (some were summarized without details in \cite{rabe:recon:17}), and some present novel contributions in themselves.
For example, to the best of our knowledge no implementation previously existed for \llfp.

We are not quite able to answer the question about the limitations of the approach: we chose our case studies in increasing order of difficulty, and they have all succeeded.
But we speculate on the limitations and future challenges in Sect.~\ref{sec:conc}.

\subparagraph*{Acknowledgments}
The rewriting case study from Sect.~\ref{sec:rewrite} was inspired by a collaboration with Gilles Dowek and the Dedukti team.
The \llfp case study from Sect.~\ref{sec:lock} was carried out with Ivan Scagnetto during the Dagstuhl Seminar 16421 on Universality of Proofs.
The linear logic case study from Sect.~\ref{sec:linear} benefited from discussions with Kaustuv Chaudhuri.
The authors were supported by DFG grant RA-18723-1 OAF and EU grant Horizon 2020 ERI 676541 OpenDreamKit.
\mmt implementation and documentation are available at \url{https://uniformal.github.io/}.
Online references for all case studies are given in the text.

%% file: syntax.tex
\begin{figure}[htb]
	\begin{center}
		\begin{tabular}{|l@{\hspace{.3cm}}l@{\hspace{.3cm}}l@{\hspace{.3cm}}l|}
			\hline
			Theory                &             &        & $T\opt{:T}=\rep{D}$ \\
			Declarations          & $D$         & $\bbc$ & $c\opt{:E}\opt{=E}\opt{\#N} \alt \rldecl{E}$ \\
			Expression            & $E$         & $\bbc$ & $x \alt \cons{c}{\Gamma}{\rep{E}}$ \\
			Context               & $\Gamma$    & $\bbc$ & $\cdot \alt \Gamma,x:E$ \\
			Notation              & $N$         & $\bbc$ & $\rep{(\vmk{n}\alt \amk{n}\alt \mathrm{string})}\;\opt{\mathtt{prec}\;\mathrm{n}}$ \\
			Number                & $n$ \\
			Identifier            & $T,c,x$\\
			\hline
		\end{tabular}
	$\begin{array}{l@{\tb\#\tb}l}
	\multicolumn{2}{l}{\texttt{LF} =} \\
	\tb\type     & \type \\
	\tb\kind     & \kind \\
  \tb\texttt{Pi}     & \{\,\vmk{1}\,\}\,\amk{1}\\
	\tb\texttt{lambda} & [\,\vmk{1}\,]\,\amk{1}\\
	\tb\texttt{apply}  & \amk{1}\,\;\,\amk{2} \\
	\tb\texttt{arrow}  & \amk{1}\to\amk{2}\\
	\end{array}$
		
\caption{Core \mmt Grammar (left) and LF Example (right)}\label{fig:grammar}
\end{center}
\end{figure}

The syntax, type system, and typing algorithms of \mmt have been described in detail in \cite{rabe:recon:17}.
In the sequel, we give a quick example-driven introduction to \mmt that builds a prototype implementation of the dependently-typed $\lambda$-calculus LF \cite{lf}.

\mmt uses no built-in symbols like $\lambda$ or $\to$.
Prototyping a formal system $S$ in \mmt means to define one \mmt theory $S$ that declares all the logical primitives of $S$ such as universes, base types, connectives, binders, inference rules, axioms, etc.
Once such a meta-theory is fixed, \mmt induces an implementation of $S$, including, e.g., parser, type reconstruction, module system, and IDE.

Fig.~\ref{fig:grammar} shows a simple fragment of \mmt's \textbf{grammar} that is sufficient for our purposes as well as an example theory for the logical primitives of LF.
An \mmt \textbf{theory} declares \textbf{constants} $c[:A][=t][\#N]$, which introduce the identifier $c$ with optional \textbf{type} $A$, \textbf{definiens} $t$, and \textbf{notation} $N$.
In a \textbf{notation}, $\vmk{i}$ indicates the position of a variable binding $x_i\!:\!A_i$, $\amk{i}$ the position of an argument $E_i$, and arbitrary strings can be used as delimiters; finally $\mathtt{prec}\, I$ determines its precedence.

$\cn{LF}$ introduces constants for the syntax of LF and their notations, e.g., the notation for \cn{lambda} introduces $\lambda$-abstractions with the concrete syntax to be $[x:A]t$.
Internally, \textbf{expressions} are represented as $\cons{c}{x_1\!\!:\!\!A_1,\ldots,x_m\!\!:\!\!A_m}{E_1,\ldots,E_n}$.
Here $c$ is the constant forming the term, the $x_i:A_i$ are variable bindings, and the $E_i$ are arguments, e.g., $[x:A]t$ corresponds to $\cons{\cn{lambda}}{x:A}{t}$, and $f\,a$ (for LF function application) to $\cons{\cn{apply}}{\cdot}{f,a}$.
Each $x_i$ is bound and may occur in $A_{i+1},\ldots,E_n$.
$\alpha$-renaming and capture-avoiding substitution are defined in the usual way.
\medskip

We can already use \cn{LF} to define other languages, e.g., as in \autoref{lst:pl1}.\footnote{from file \url{https://gl.mathhub.info/MMT/examples/blob/devel/source/logic/pl.mmt}}
Here, $PL:LF=\Sigma$ defines the theory \cn{PL} (for propositional logic) with \textbf{meta-theory} LF, and body $\Sigma$.
Apart from LF being an \mmt theory itself, this definition of propositional logic in LF is standard.
Note how the constants \cn{andI} and \cn{impI} (for proof rules) have notations that do not list all their arguments --- the other arguments are implicit and have to be reconstructed.
Similarly, omitted types of bound variables are reconstructed.
\medskip

\begin{mmtcode}[caption={Fragment of an MMT Theory for propositional logic},label=lst:pl1]
theory PL : LF =
	prop  : type
	ded   : prop ⟶ type         # ⊦ 1  prec -5
	and   : prop ⟶ prop ⟶ prop   # 1 ∧ 2 prec 15
	impl  : prop ⟶ prop ⟶ prop   # 1 ⇒ 2 prec 10
	equiv : prop ⟶ prop ⟶ prop   # 1 ⇔ 2 prec 10
           = [x,y] (x ⇒ y) ∧ (y ⇒ x)
	andI  : {A,B} ⊦ A ⟶ ⊦ B ⟶ ⊦ A ∧ B   # andI 3 4
	implI: {A,B} (⊦ A ⟶ ⊦ B) ⟶ ⊦ A ⇒ B  # impI 3
        
        equivI : {A,B} (⊦ A ⟶ ⊦ B) ⟶ (⊦ B ⟶ ⊦ A) ⟶ ⊦ A ⇔ B  
           = [A,B,p,q] andI (impI [x] p x) (impI [x] q x)
\end{mmtcode}

In $T:M=\Sigma$, the purpose of the \textbf{meta-theory} $M$ is to induce the syntax and semantics of $\Sigma$.
All constants and notations of $M$ are available to form the types and definitions of the constants in $\Sigma$.
And the rules of $M$ define how $\Sigma$ is type-checked.
To understand how to supply rules for \cn{LF}, we need a little more background about the \mmt type system. 
\medskip

\begin{figure}[ht]\centering
	\begin{tabular}{|l|l|} 
		\hline
		Judgment & Intuition \\ \hline
		\multicolumn{2}{|l|}{All judgments are relative to fixed theory $\Theta$ and its meta-theories.}\\
		\hline
		\hline $\gjudg{\judginh T}$ & $T$ is inhabitable (may occur as the type of a constant) \\
		\hline\hline $\gjudg{ \hastype t T}$ & $t$ checks against inhabitable term $T$. \\
		\hline $\gjudg{ \infertype t T}$ & the principal type of term $t$ is inferred to be $T$ \\
		\hline $\gjudg{ \judgeq {t_1} {t_2} T}$ & $t_1$ and $t_2$ are equal at type $T$ \\
		\hline $\gjudg{ \judgeqc {t_1} {t_2}}$ & $t_1$ computes to $t_2$ (preserving type) \\
		\hline
	\end{tabular}
	\caption{Judgments}\label{fig:judgments}
\end{figure}

The \textbf{judgments} are given in \autoref{fig:judgments}.
For simplicity, we assume that a set of theory definitions has been fixed and that all judgments are relative to one theory $\Theta$, without making this explicit in the notation.
No constant-specific rules are built in:
the only rules fixed by \mmt are lookup rules (e.g., to infer the type of a variable $x$), equivalence and congruence of equality, and $\alpha$-renaming of bound variables.
The four judgments for equality and typing are standard for bidirectional type-checking.
The unusual judgment $\gjudg{\judginh T}$ is used to check constant declarations: the declaration $c:T$ is allowed if $\gjudg{\judginh T}$, and the declaration $c:T=t$ additionally requires $\gjudg{\hastype t T}$.
\medskip

Finally, we add \textbf{rule declarations} $\rldecl{E}$ to the theory \cn{LF}.
Here the expression $E$ is interpreted as an object of the underlying programming language (Scala).
When checking a judgment, \mmt uses only those rules that are visible to the current context.
Each rule is a self-contained Scala object that is loaded dynamically when needed.
When prototyping systems in \mmt, users usually use a logical framework like LF and only define declarative, statically checked rules like we did for \cn{PL} above.
But, critically, users have the option to add a Scala-based rule whenever necessary.
\medskip

Each rule implements a specific Scala interface.
For example, rules implementing \emph{InferenceRule} are used whenever the judgment $\infertype t T$ is encountered; this interface includes a method \emph{inferType} that receives the current context and $t$ and must try to return $T$.
It also receives callbacks to recursively check the premises of the rule. \autoref{lst:lambdascala} shows an example rule inferring the type of a lambda term.\footnote{in file \url{https://github.com/UniFormal/MMT/blob/devel/src/mmt-lf/src/info/kwarc/mmt/lf/Rules.scala}}
\begin{scalacode}[caption={A Rule Inferring the Type of a Lambda Term},label=lst:lambdascala,emph={Lambda,Pi}]
object LambdaTerm extends InferenceRule(Lambda.path) {
   def apply(solver: Solver)(tm: Term, covered: Boolean)(implicit stack: Stack, history: History)
    : Option[Term] = tm match {
        case Lambda(x,a,t) =>
           if (!covered) isTypeLike(solver,a)
           val (xn,sub) = Common.pickFresh(solver, x)
           solver.inferType(t ^? sub, covered)(stack ++ xn 
        case _ => None // should be impossible
      }
   }
\end{scalacode}

For LF, we use $5$ inference rules applicable to different terms: one each for terms formed from $\type$, $\kind$, \cn{Pi} (or \cn{arrow}), \cn{lambda}, and \cn{apply}.
Additionally, we define $2$ rules that check $\hastype t T$ and $\judgeq t {t'} T$ whenever $T$ is a \cn{Pi}-expression, and a computation rule for $\beta$-reduction.
For type-reconstruction, \mmt uses some variables to represent omitted subexpressions that must be reconstructed.
For LF, only one additional rule is needed that is aware of these meta-variables --- a special rule for pattern unification.

Together with two rules for inhabitability, our LF prototype requires $11$ Scala rules, taking about $200$ lines of Scala code.
This is a tiny fraction compared to the logic-independent code in \mmt that builds a full-fledged application from these $11$ rules.
Moreover, these $200$ lines that carry the semantics of LF are all in one place and much easier to read and verify than a from-scratch implementation.

%% file: dhol.tex
\begin{framed}
	By adding a single new typing rule to LF, we enable shallow polymorphism.
	This allows developing dependently-typed higher-order logic.
\end{framed}

LF users have often lamented the lack of polymorphism to define, e.g., polymorphic equality, lists, product types, etc.
An important observation is that shallow polymorphism, i.e., the binding of kinded variables but only at toplevel, is much simpler than full polymorphism but already enables many important applications.

To enable shallow polymorphism, we extend our LF prototype with a single rule\footnote{in file \url{https://github.com/UniFormal/MMT/blob/devel/src/mmt-lf/src/info/kwarc/mmt/lf/ShallowPolymorphism.scala}} for the judgment $\judginh{\{x:A\}B}$.
This rule infers the type of $A$, checks that it is a universe (i.e., $\type$ or $\kind$) and then recursively checks that $B$ is inhabitable.
Because the judgment $\judginh T$ is only called at toplevel, this yields exactly the desired effect.
Maybe surprisingly, except for a few minor adaptations, all existing rules of \cn{LF} can remain unchanged, e.g., \cn{LF}'s Scala rule to check $\hastype t{\{x:A\}B}$ worked immediately if $A$ is a kind.  

We call the resulting logical framework PLF.
Using the \mmt module system, it is defined by including \cn{LF} and adding one rule declaration.\footnote{in file \url{https://gl.mathhub.info/MMT/urtheories/blob/devel/source/primitive_types/bool.mmt}}.

%
%
%

The resulting case study is shown in \autoref{lst:dhol}.
Starting with polymorphic equality, we can define higher-order logic in the usual way, including the definition of polymorphic quantifiers and their proof rules.
Note how kinded variables are treated in the same way as typed ones because \mmt does not care about the difference anyway.
Similarly, type reconstruction of variable kinds and implicit type arguments works out of the box: the new Scala rule does not even mention reconstruction.

All declarations are straightforward except for the congruence rule \cn{cong}.
We stated it here for simply-typed functions only due to the inherent difficulties of combining dependent types with equality.
But this is a well-known theoretical problem and not an \mmt issue.
In fact, \mmt can now help with the problem: It is often unclear how well a potential solution for this deep problem works in practice.
By building variants of \cn{DHOL}, possibly with additional Scala rules to handle equality specially, we can quickly prototype and evaluate possible solutions.

\begin{mmtcode}[caption={An MMT Theory for Dependently-Typed HOL},label=lst:dhol]
theory DHOL : PLF =
  bool  : type 
  equal : {A:type} A ⟶ A ⟶ bool  # 2 ≐ 3 prec 5 
  ded  : bool ⟶ type  # ⊦ 1 prec -5 
  refl : {A,X:A} ⊦ X ≐ X  # refl 
  cong : {A, B: type} {F: A ⟶ B} {X, Y: A} ⊦ X ≐ Y ⟶ ⊦ (F X) ≐ (F Y)  # cong 3 6 

  extensionality : {A:type,B:A ⟶ type}{F:{x:A} B x, G:{x:A} B x} ({x: A} ⊦ F x ≐ G x) ⟶ ⊦ F ≐ G 
    # ext 5 
  boolEqIntro : {B1,B2:bool} (⊦ B1 ⟶ ⊦ B2) ⟶ (⊦ B2 ⟶ ⊦ B1) ⟶ ⊦ (B1 ≐ B2)  # beqI 3 4
  boolEqElim : {B1,B2:bool} ⊦ B1 ≐ B2 ⟶ ⊦ B1 ⟶ ⊦ B2  # beqE 3 4 
  symmetry : {A : type}{a,b : A} ⊦ a ≐ b ⟶ ⊦ b ≐ a 
    = [A][a,b][p] (cong ([x] x ≐ a) p) refl  # symm 4 
  eqFun : {A : type,B : type}{F,G : A ⟶ B} ⊦ F ≐ G ⟶ {a} ⊦ F a ≐ G a 
    = [A,B][F,G][p][a] cong ([H: A ⟶ B] H a) p # eqfun 5 6 

  true  : bool  = ([x:bool] x) ≐ ([x:bool] x) 
  trueI : ⊦ true  = refl 
  forall : {A:type} (A ⟶ bool) ⟶ bool  = [A,P] P ≐ ([x] true)  # ∀ 2 
  forallI : {A:type, P : A ⟶ bool}({x} ⊦ P x) ⟶ ⊦ ∀[x] P x 
    = [A,P][p] ext ([x: A] beqI ([pf: ⊦ P x] trueI) ([pt: ⊦ true] p x)) 
  forallE : {A:type, P : A ⟶ bool} (⊦ ∀[x]P x) ⟶ {a} ⊦ P a 
    = [A,P][p][a] beqE (eqfun (symm p) a) trueI 
\end{mmtcode}

%% file: rewriting.tex
\begin{framed}
	We build an MMT plugin that interprets certain declarations as rewrite rules and adds them on the fly.
	This yields LF modulo rewriting akin to Dedukti.
\end{framed}

LF modulo rewriting \cite{lambdaPimodulo} as implemented in Dedukti \cite{dedukti} was recently identified as a sweet spot in logical framework design --- a simple feature with huge practical benefit.
Inspired by this, we want to extend our LF prototype to LF modulo.

In LF, a rewrite rule can be seen as an LF constant $c:\{\Gamma\}l\doteq r$, which rewrites $l$ to $r$ in context $\Gamma$.
(For simplicity, we reuse the equality predicate $\doteq$ from Section~\ref{sec:dhol}.)
In plain LF, this is just a normal constant that has no bearing on computation.
In LF modulo, this rule must result in an additional computation rule $\gjudg{\judgeqc l r}$.

This naturally leads to the main idea of our prototype study: we build an \mmt plugin that automatically adds a rule declaration when a constant with an appropriate type is encountered.
For this we use some machinery that is available to \mmt plugins but that we have not mentioned in Section \ref{sec:prelim}.

Firstly, plugins can listen to \mmt events.
Whenever a constant $c$ is added, our plugin inspects it and possibly adds a new computation rule declaration to \mmt's internal data structures.
The rule declaration is added right after the declaration of $c$ so that it is in scope whenever $c$ is.

Secondly, like in Dedukti, we do not check confluence or termination so that the user needs fine-grained control over which declarations induce rewrite rules.
For that, we exploit that every \mmt declaration may carry metadata.
Most importantly, it provides convenient syntax to attach a \emph{role} to each constant.
The role has no effect unless plugins choose to act on it.
Our plugin picks up on the role \cn{Simplify} --- any constant with this role becomes a rewrite rule.
\medskip

The implementation of our plugin -- including reusable abstractions for matching -- takes about 450 lines of code\footnote{in file \url{https://github.com/UniFormal/MMT/blob/devel/src/mmt-lf/src/info/kwarc/mmt/lf/SimplificationRuleGenerator.scala}}.
To evaluate it, we formalize $\Sigma$-types.
In general, there are three different ways to do that:
\begin{itemize}
	\item As new primitive features akin to how we defined $\Pi$-types in \cn{LF}.
	But that requires new Scala rules.
	\item By declaring an object logic inside LF. But that does not allow building pairs of arbitrary LF types.
	\item By defining polymorphic $\Sigma$-types in \cn{PLF}.
\end{itemize}
All three ways have their merit, and the latter two can make use of rewriting in essentially the same way: they turn the two $\beta$-style rules into rewrite rules $\judgeqc{\cn{pi}_i(\cn{pair}\,x_1 \,x_2)}{x_i}$.
Here we choose the third way, both for the sake of example and because it is a neat trade-off between the other two options.
The resulting formalization is given in \autoref{lst:sigma1}\footnote{in file \url{https://gl.mathhub.info/MMT/examples/blob/devel/source/sigma.mmt}}.

\begin{mmtcode}[caption={$\Sigma$-types defined in dependently-typed HOL},label=lst:sigma1]
theory Sigma : DHOL =
    Sigma : {A:type} (A ⟶ type) ⟶ type # Σ 2 
    pair : {A : type, B: A ⟶ type, a : A } B a ⟶ Σ B # pair 3 4 
    pi1 : {A : type, B: A ⟶ type} Σ B ⟶ A # pi1 3 
    pi2 : {A : type, B: A ⟶ type} {p : Σ B} B (pi1 p) # pi2 3

    conv_pair : {A: type, B: A ⟶ type, u: Σ B} ⊦ u ≐ (pair (pi1 u) (pi2 u))
    conv_pi1 : {A : type, B: A ⟶ type}{a : A, b: B a} ⊦ (pi1 (pair A B a b)) ≐ a	role Simplify
    conv_pi2 : {A : type, B: A ⟶ type}{a : A, b: B a} ⊦ (pi2 (pair A B a b)) ≐ b	role Simplify
\end{mmtcode}

All declarations are straightforward except for \cn{conv\_pi2}.
In plain LF, this declaration does not type-check: the two sides of the equality have different types, namely \texttt{B (pi1 (pair A B a b))} on the left and \texttt{B b} on the right.
But in LF modulo, it type-checks because \cn{conv\_pi1} is registered as a rewrite rule.

The combination of shallow polymorphism and rewriting is an amazingly powerful extension while retaining the look-and-feel of LF.
For example, it allows defining polymorphic lists with their induction principles and rewrite rules to compute recursive functions.
The separation of concerns that \mmt introduces allowed us to build this extension of LF within days without changing any line in \mmt itself.

%% file: locks.tex
The previous case studies considered well-understood extensions of LF, which makes prototyping simpler.
Therefore, the next case study considers an experimental formal system that was introduced only recently, as a theoretical result without mature implementation.
Such new systems abound, and it is here where rapid prototyping is most urgently needed.

\begin{framed}
	Lock types were introduced in \cite{llfp} to allow for arbitrary external side conditions in LF.
	Using MMT, we obtain a prototype implementation within a few hours.
\end{framed}

\newcommand{\Lp}{\mathcal{L}}

The central idea of \llfp is to extend LF with a monadic type operator $\Lp^K$ that is parametrized by an arbitrary extra-linguistic side condition $K$, which is of the form $p(t,A)$ for a term $\hastype t A$ and an identifier $p$.
Within the monad, $K$ can be assumed at no cost.
But to access a monadic value from the outside, the side condition has to be evaluated.

Our case study represents \llfp and applies it to one of the examples described in \cite{llfp} --- a call-by-value reduction of the $\lambda_v$ calculus.
Here the side condition is of the form $K=\cn{Val}(t,\cn{term})$, and $\cn{Val}$ checks that $t$ is a $\lambda_v$-value (i.e., a variable or a $\lambda$-abstraction).
Such syntactic side conditions are quite common but cannot be formalized in LF because they are not preserved by substitution.

We formalize the syntax of \llfp in theory \cn{LLFP} in \autoref{lst:lockmmt}: we include \cn{LF} and extend it with one constant each for the formation, introduction, and elimination form of the new type constructor. 
We only need to write $6$ rules in Scala that require roughly $100$ lines of code.
Additionally, we introduce a new abstract rule interface *ExternalConditionRule* for rules that check our side conditions.
*ExternalConditionRule* is a new abstract interface, and such rules can evaluate $K$ in any way they want.
Their behavior is as follows:
(i) The formation rule infers $\infertype{\Lp\,K\,\langle A\rangle}{\type}$ from $\infertype{A}{\type}$.
(ii) The introduction rule infers $\infertype{L\,K\,\langle t\rangle}{\Lp\,K\,\langle A\rangle}$ from $\infertype{t}{A}$.
Both rules use the flexibility of \mmt expressions to put the assumption $K$ into the context --- something that plain LF would not allow because $K$ is not a well-typed LF expression.
(iii) The elimination rule infers $U\langle t\rangle:A$ if $t:\Lp\,K\,\langle A\rangle$ and if $K$ holds.
This rule is called in *InferUnlock* in Scala, and a fragment of its code is given in Figure~\ref{lst:llfpscala}.
It establishes $K$ of the form $p(t,A)$ in one of two ways: if $K$ is in the context (i.e., we are in the monad), it is assumed; otherwise, we look up an *ExternalConditionRule* for $p$ and call it on $K$.
(iv-vi) The remaining $3$ rules (check against monadic type, compare equality at monadic type, reduce unlock of lock) are straightforward.

The $\lambda_v$ case study is shown in the bottom part of \autoref{lst:lockmmt}\footnote{in file \url{https://gl.mathhub.info/MMT/urtheories/blob/devel/source/llfp.mmt}}.
We declare the side condition \cn{Val} as an untyped constant and a rule \cn{ValRule} of type *ExternalConditionRule*.
The Scala code of \cn{ValRule} is given in \autoref{lst:llfpscala}\footnote{in file \url{https://github.com/UniFormal/MMT/blob/devel/src/mmt-lf/src/info/kwarc/mmt/lf/externals/ExternalCheck.scala}}.
It checks the condition $\cn{Val}(t,a)$, i.e., that $t$ is of the form $\cn{free}\,n$ or $\lambda\, F$; if that fails, it registers an error message that the IDE will show to the user.

The remainder then follows \cite{llfp} in a straightforward way.
We added the declaration \cn{check2} and \cn{fail}, where *InferUnlock* calls *ValRule*.
This succeeds in the first case because the argument \texttt{'n}, which appears as *LF.Apply(CallByValue.free, n)* in the Scala code, is a $\lambda_v$-value; it (correctly) fails in the second case where the argument is not a $\lambda_v$-value.

\begin{mmtcode}[caption={An MMT Theory for the Syntax of Lock Types},label=lst:lockmmt]
theory LLFP =
   include LF
   locktype # $\Lp$ 1 $\langle$ 2 $\rangle$ // type formation: $\Lp K$ is the monadic operator
   lockterm # L 1 $\langle$ 2 $\rangle$ // introduction form
   unlock   # U $\langle$ 1 $\rangle$   // elimination form
   // 6 rules omitted   

theory CallByValueExample : ?LLFP =
   term : type
   app : term ⟶ term ⟶ term  # 1 @ 2 prec 50
   lam : (term ⟶ term) ⟶ term # λ 1
   ... // (natural numbers omitted)
   free : nat ⟶ term # ' 1 prec 100
   
   eq : term ⟶ term ⟶ type  # 1 ≐ 2
   eq_app: {M,N,X,Y} M ≐ N ⟶ X ≐ Y ⟶ M@X ≐ N@Y  # eq_app 5 6
   ... // (additional equality judgment and its rules omitted)
   
   Val
   rule rules?ValRule
   
   // reduction rules using Val condition
   betav : {M,N} $\Lp$ (Val N term) $\langle$(λ M)@N≐(M N)$\rangle$ # betav 1 2
   csiv : {M,N}({x} $\Lp$ (Val x term) $\langle$(M x)≐(N x)$\rangle$) ⟶ (λ M) ≐ (λ N) # csiv 3
   
   // example from the end of the section
   goal = λ[x]'0 @ ((λ[y] y) @ x) ≐ λ[x] '0 @ x

   // The following check succeeds without ever calling the ValRule because the unlock is under a lock.
   check: goal 
     = csiv [x] L (Val x term) $\langle$
       eq_app (refl '0) U$\langle$betav ([y] y) x$\rangle$
     $\rangle$
   // The following example does not guard the unlock but still succeeds because 'n is indeed a value.
   check2 = [n] eq_app (refl '0) U$\langle$betav ([y] y) 'n$\rangle$
   // The following negative example fails because x is any term and thus not necessarily a value.
   fail   = [x] eq_app (refl '0) U$\langle$betav ([y] y) x$\rangle$
\end{mmtcode}

\begin{scalacode}[caption={A Rule Inferring the Type of a Lock Term},label=lst:llfpscala,emph={LockTerm,Key,LockType}]
object InferUnlock extends InferenceRule(Unlock.path) {
	def apply(solver: Solver)(tm: Term)(context: Context) : Option[Term] = {
		... // side condition is p(t,a)
		solver.rules.getByHead(classOf[ExternalConditionRule], p).headOption match {
			case Some(rule) =>
			   if (! rule(solver)(context, t, a))
			     return None // side condition failed
			case None =>
			   solver.error("no rule for condition " + p + " found")
		}
		...
	}

object ValRule extends ExternalConditionRule(CallByValue.Val) {
  def apply(solver: Solver)(context: Context, t: Term, a: Term) : Boolean = {
    t match {
      case LF.Apply(CallByValue.free, _) => true
      case LF.Apply(CallByValue.lam, _) => true
      case _ => solver.error(solver.presentObj(t) + " must be lambda-abstraction or variable")
    }
  }
}
\end{scalacode}

During this case study, we found a few possible simplifications to the \llfp calculus --- minor tweaks that make the language and implementation easier but do not appear to change its essence.
That shows rapid prototyping at its best: with MMT's parser, type reconstruction, and IDE working out of the box, we could tweak the \llfp grammar and update the rules and the $\lambda_v$ example with just a few minutes of turnaround time.
That was critical to conduct the experimentation needed to find these simplifications.

%% file: linear.tex
The previous case studies focused on systems relatively near to LF, for which \mmt was originally designed.
The next case study takes us out of this comfort zone by considering a substructural logic, which tends to be incompatible with LF.

\begin{framed}
	Linear logic is tricky to represent in standard (i.e., not substructural) frameworks.
    The resource semantics provides an appealing solution, but it is only elegant if a free monoid is available.
    In \mmt, we can simply add one by writing appropriate equality rules.
\end{framed}

Substructural frameworks omit one or several of the basic principles \emph{exchange}, \emph{weakening} or \emph{contraction}. Since LF uses all of them, this makes elegant representations of substructural logics difficult.
The resource semantics of linear logic interprets formulas in a Kripke frame, modeling the linear contexts as worlds.
Consequently, this requires a free commutative monoid structure for the type of worlds.
Our formalization follows Frank Pfenning's lecture notes.

\begin{mmtcode}[caption={Resource Semantics of Linear Logic},label=lst:linear1]
theory ResourceSemantics : LF =
	o : type
	⊗ : o ⟶ o ⟶ o # 1 ⊗ 2
        ⊸ : o ⟶ o ⟶ o  # 1 ⊸ 2 prec 80
	... // other connectives omitted
	
	// worlds contain resources 
	world : type
	// composition of worlds, corresponds to multiset union of resources 
	union : world ⟶ world ⟶ world  # 1 * 2 prec 100
	// empty world with no resources 
	empty : world # ε
	// truth at a world
	At : o ⟶ world ⟶ type # 1 @ 2 prec 50
	// validity is truth in the empty world
	valid : o ⟶ type = [A] A @ ε

	... // axioms for commutative monoid omitted
	
	// proofs of the linear sequent A1, ..., An |- A are represented as terms of type
	// (A1 ⊗ ... ⊗ An) ⊸ A  @  ε, where @ is the binary holds-at relation between 
	// propositions and worlds
	⊗_R  : {A,B,a,b} A @ a ⟶ B @ b ⟶ A ⊗ B @ a*b
	⊗_L  : {A,B,C,u,v} A ⊗ B @ u ⟶ ({a}{b} A @ a ⟶ B @ b ⟶ C @ a*b*v) ⟶ C @ u*v
	# ⊗_L 6 7
	⊸_R  : {A,B,w} ({a} A @ a ⟶ B @ a*w) ⟶ A ⊸ B @ w
	... // other rules omitted
\end{mmtcode}

We give a fragment of our formalization in \autoref{lst:linear1}\footnote{see file \url{https://gl.mathhub.info/MMT/examples/blob/devel/source/logic/linear.mmt}}.
It uses plain LF as the meta-theory and declares the linear propositions, the free commutative monoid of worlds, the holds-at relation between propositions and worlds, and the proof rules.

This formalization has the problem that proofs in linear logic now require equational reasoning about the worlds.
The resulting proofs become inelegant and cumbersome.
Therefore, we add two rules\footnote{in file \url{https://gl.mathhub.info/MMT/examples/blob/devel/scala/info/kwarc/mmt/examples/ResourceSemantics.scala}} that automate some of the equational reasoning.
This is undecidable. But in linear logic proofs, the worlds are usually formed from bound variables only so that we can find reasonably good heuristics:
(i) The rule *NormalizeWorlds* for $\judgeqc v w$ at type \texttt{world} removes brackets (associativity) and units (neutrality) and normalizes the order of bound variables (commutativity).
(ii) The rule *EquateWorlds* for judgment $\judgeq v w {\cn{world}}$ normalizes both $v$ and $w$ and then tries to equate them.
Here we make use of the freeness property: in the free monoid, if $v,v',w$ are bound variables, $\judgeq{v\ast w}{v'\ast w}{\cn{world}}$ is equivalent to $\judgeq v {v'}{\cn{world}}$.
This is important to make type reconstruction recurse into equality checks that eventually solve meta-variables of type \cn{world}.
This is necessary to make the world parameters of the proof rules implicit arguments, which in turn is necessary to write proofs in practice.
Both rules are straightforward to implement and take just a few dozen lines of Scala code.

If we add these rules, we can write linear logic proofs elegantly.
An example is given in \autoref{lst:linear4}.

\begin{mmtcode}[caption={A Proof for Associativity of $\otimes$},label=lst:linear4]
   // reconstruction must solve abc * X = abc and ab * Y * X = ab * c with X = epsilon, Y = c
   ⊗_assoc1 : {A,B,C} valid (A⊗B)⊗C ⊸ A⊗(B⊗C)
        = [A,B,C] ⊸_R [abc,h] ⊗_L h 
              [ab,c,pq,r] ⊗_L pq
                 ([a,b,p,q] ⊗_R p (⊗_R q r))
\end{mmtcode}

Prototypes of other substructural logics can be obtained accordingly, e.g., by using idempotence to model the contraction rule.

%% file: hott.tex
The previous case studies considered relatively simple formal systems that use very few language features and thus require only a few rules.
Therefore, the next case study considers a more complex system combining several typing features.

\begin{framed}
	Homotopy Type Theory (HoTT,\cite{hottbook}) is a foundation of mathematics based on a dependently typed Martin-Löf type theory and the univalence axiom.
	We implement it in MMT using $\approx 45$ rules.
\end{framed}

HoTT is a relatively young formal system that has received significant attention and multiple variants are actively investigated.
That makes it a prime candidate for rapid prototyping.

HoTT uses $\Pi$-types, $\Sigma$-types, coproduct types, finite types, W-types (which represent inductive types), a countable cumulative hierarchy of type universes, and intentionally proof-relevant equality types.
We formalize each feature as a theory in the same way in which we formalized $\Pi$-types in \cn{LF} and eventually obtain our HoTT prototype as the union of all theories.

We only present our definition of W-types here because it is the most unusual from the \mmt perspective.
The type $\wtype{x:A}B(x)$ represents the inductive type whose constructors are the terms $x:A$ whose arity is represented by $B(x)$.
Typically, types $A$ and $B(x)$ are finite.
For example, the natural numbers correspond to $\wtype{x:\texttt{Bool}}{x\texttt{\ match}\{\texttt{true}\Rightarrow\varnothing,\;\texttt{false}\Rightarrow\texttt{Unit}\}}$.

The elimination rule, for example, is:
{\small For $C':={C\sub c{\wsupu cx{g(x)}{\wtype{x:A}B}}}$:}
\[
\trule{\gjudg{\infertype w{\wtype{x:A}B}} \quad \judg{\Gamma,c:\wtype{x:A}B}{\infertype{C}U} \quad \judg{\Gamma,c:A,g:\lfarrow{B\sub xc}{\wtype{x:A}B},h:\lfpi{y:B\sub xc}{C\sub c{g(y)}}}{\hastype e{C'}} }{\gjudg{\infertype{\wrec {\wtype{x:A}B}wCcghe}{C\sub cw}}}
\]

$\wtypesym$-types are a relatively simple representation of induction in a type-theoretical setting and are well-understood from an implementation point of view.
Indeed, implementing the Scala rules is tedious but straightforward.
But from a user perspective, even formalizing seemingly simple inductive types like natural numbers or lists using $\wtypesym$-types is inconvenient and unintuitive.
For example, \autoref{lst:natugly}\footnote{in file \url{https://gl.mathhub.info/MMT/LFX/blob/devel/source/test.mmt}} shows an implementation of natural numbers and addition.

\begin{mmtcode}[caption={Natural Numbers as $\wtypesym$-types},label=lst:natugly]
theory Wtest :?LFW =
	beta : ENUM 2 ⟶ type  = [x] x match y. ∅ |UNIT to type 
	ℕ : type  = W x:ENUM 2 . (beta x) 
	
	zeroN : ℕ  = sup (CASE 0) , x ⟹ (x ∅f ℕ) to ℕ  
	S : ℕ ⟶ ℕ  = [x : ℕ] sup (CASE 1) , y ⟹ x to ℕ 
	
	plusN : {n:ℕ,m:ℕ} ℕ  = [n][m] (rec m,c,g,h ⟹ (c match  (ENUM 2), x. n | S (h x) to ℕ) to ℕ) 
	   # 1 + 2 prec 5 
\end{mmtcode}

In order to provide more convenient syntax, we write a plugin that extends the \mmt language.
This uses a powerful extension mechanism of \mmt --- called \emph{structural features} --- that we omitted in Figure~\ref{fig:grammar} for simplicity.
It is described in \cite{Iancu:phd}.
Our structural features are intended to enable the definitions shown in \autoref{lst:natnice}\footnote{same file}, which is immediately recognizable as an inductive type and a recursive function.

\begin{mmtcode}[caption={Natural Numbers using Structural Features},label=lst:natnice]
theory FeatureTest : LFX/WTypes?Inductive =
	induct Nats () =
		Nat : type  # ℕ
		Zero : ℕ  # O 
		Succ : ℕ ⟶ ℕ  # S 1 
	def addition (n : ℕ)  =
		add : ℕ ⟶ ℕ  # 1 + 2 
		Zero = n 
		Succ = [m] S (add m) 
\end{mmtcode}

Structural features can be seen as an elegant logic-independent version of Isabelle's definitional packages.
Syntactically, they allow using arbitrary nested theories as declarations.
Semantically, these nested theories are elaborated into a list of declarations generated from its contents.
Each such nested theory is introduced by a special keyword that indicates which structural feature provides the elaboration function.

For $\wtypesym$-types, we provide two structural features for the keywords \texttt{induct} and \texttt{def}.
Both are conservative in the sense that the elaboration contains only \emph{defined} constants.
The former generates a constant with a $\wtypesym$-type as definiens and analogously defined constants for all constructors of the type.
The latter generates an inductively defined function on a $\wtypesym$-type using the cases given in the body.
For example, \autoref{lst:feature1}\footnote{ in file \url{https://gl.mathhub.info/MMT/LFX/tree/devel/scala/info/kwarc/mmt/LFX/WTypes/Features.scala}} shows the relevant part of code that generates the $\wtypesym$-type.

\begin{scalacode}[caption={Code that elaborates the Body of a Theory into a $\wtypesym$-type},label=lst:feature1]
implicit val tpname : (GlobalName,Term) = (tpc.path,if (params.isEmpty) 
     OMS(parent.path.module ? tpc.name) else
      ApplySpine(OMS(parent.path.module ? tpc.name),params.variables.map(_.toTerm):_*))
val cases = consts.tail.zipWithIndex.map(p => makeCase(p._1,p._2))

val tpA = Coprod(cases.map(_.getcptype):_*)
val (xn,_) = Context.pickFresh(dd.getInnerContext,LocalName("x"))
val (yn,_) = Context.pickFresh(dd.getInnerContext,LocalName("y"))
val tpB = cmatch(OMV(xn),tpA,yn,cases.map(_.getArity),OMS(Typed.ktype))
val tp = if (params.isEmpty) Some(OMS(Typed.ktype)) else Some(Pi(params,OMS(Typed.ktype)))
val df = if (params.isEmpty) Some(WType(xn,tpA,tpB)) else Some(Lambda(params,WType(xn,tpA,tpB)))
Some(Constant(parent.toTerm,tpname._1.name,tpc.alias,tp,df,tpc.rl,tpc.notC))
\end{scalacode}

Assuming we have collected all parts of HoTT in a theory \cn{Types}, we can finish the HoTT prototype by stating the univalence axiom.
This is shown in \autoref{lst:hott}\footnote{in file \url{https://gl.mathhub.info/MMT/LFX/tree/devel/source/HOTT.mmt}}.
Here \texttt{NAT} is the type of natural numbers, and $\mathcal{U}$ forms the universes. 

\begin{mmtcode}[caption={A Theory for HoTT},label=lst:hott]
theory HOTT : Types =
	
	identity_fun : {i: NAT, A : 𝒰 i} A ⟶ A  = [i,A]([x : A] x)  # id 2 
	
	homotopy : {i : NAT}{A : 𝒰 i, B : A ⟶ 𝒰 i} ({x:A} B x) ⟶ ({x:A} B x) ⟶ 𝒰 i 
		= [i][A, B][f][g] {x:A} (f x ≐ g x)  # 4 ∼ 5 
			
	is_equiv : {i : NAT}{A:𝒰 i,B:𝒰 i} (A ⟶ B) ⟶ 𝒰 i 
		= [i][A,B][f] (Σ g : B ⟶ A . ([x:B] f (g x)) ∼ (id B)) × 
			(Σ h : B ⟶ A . ([x:A] h (f x)) ∼ (id A))  #  isequiv 4 
	  	
	equivalence : {i : NAT}𝒰 i ⟶ 𝒰 i ⟶ 𝒰 i 
	  	= [i][A,B] Σ f : A ⟶ B . isequiv f  # 2 ≃ 3 
	  
	univalence : {i : NAT}{A:𝒰 i,B:𝒰 i} (A ≃ B) ≃ (A ≐ B) 
\end{mmtcode}

Our definition of HoTT followed the original one from \cite{hottbook}.
Since then, a flurry of activity has led to many different variants, in particular regarding the treatment of equality and the formation of higher inductive types.
Our prototype provides an ideal starting point to experiment with these variants.

%% file: conc.tex
\subparagraph*{Successes}
We have evaluated the \mmt framework from a rapid prototyping perspective by defining $5$ formal systems in it.
Our experiences highlight two essential strengths:
\mmt is very flexible thus allowing for elegant representation of diverse systems --- critical because prototypes are often needed exactly for novel ideas to which existing framework are not applicable.
And it makes the process of implementing and testing these representations fast and convenient --- critical because it enables a rapid feedback cycle between theory and practice.

These findings are supported by other case studies that we did not mention here.
Major previously published case studies defined record types (including defined fields and the reflection of theories into records) \cite{MueRabKoh:tat18} and the module system of PVS \cite{KMOR:pvs:17}.
Unpublished results include LISP-style quasi-quotation, sequence arguments, and intersection types, more general inductive types (Colin Rothgang), and the diagram presentation language of  \cite{mathschemetheoexp} (Yasmine Sharoda).

We are also working on several exports of proof assistant libraries relative to representations of Isabelle/Pure (with Makarius Wenzel) and Coq (with Claudio Sacerdoti Coen) in \mmt.
In fact, such exports were part of the original motivation of \mmt:
Representations of proof assistant libraries in standardized frameworks had previously suffered because existing frameworks were too inflexible to represent the respective logics in ways that naturally match the implementations in the proof assistants.

To further evaluate the approach, increasingly challenging benchmarks should be attempted.
We expect this will be successful for, e.g., Coq-style type classes and unification hints, Idris-style meta-programming, Beluga-style reasoning about contexts, or Abella-style two-level logic and $\nabla$ operator.

\subparagraph*{Limitations}
There are also challenges that, while possible, are currently somewhat inconvenient.
While we are confident that these do not constitute limitations of the overall approach, they show limitations in the current system.
Most importantly, \mmt needs support for proof search in order to prototype formal systems with undecidable typing relations, where type reconstruction must synthesize a proof term.
Examples are predicate subtypes and quotient types, undefinedness, and Mizar-style soft type systems.
While proof search is already possible in \mmt, it currently suffers from the lack of an efficient prover.

We prioritized writing declarative rules in logical frameworks in order to minimize the amount of user-written Scala code that has to be trusted.
In particular, we used a plugin to operationalize declarative specifications by generating rewrite rules in Section~\ref{sec:rewrite}.
But we could still do better by dynamically generating more complex rules from appropriately tagged declarations.
For example, we could generate typing rules to handle unification hints or soft typing.
It is unclear if this is already possible or requires adaptations of the \mmt kernel.

Finally, \mmt has prioritized making the \emph{prototyping} fast rather than the \emph{prototype}:
it focuses on making prototypes usable (flexible notations, error messages, IDE, etc.) and just fast enough for interactive case studies.
While \mmt rules are fast because they are written in a mainstream programming language, \mmt loses speed by explicitly dispatching to these rules (rather than using the dispatch mechanism of the underlying programming language).
Other systems like ELPI \cite{elpi} or Dedukti \cite{dedukti} have instead optimized for the speed of the prototype.
While the speed of the prototypes was not a bottleneck in our case studies, it could become problematic if \mmt-based prototypes are evaluated on large inputs, e.g., for interactive theorem proving (as in ELPI) or batch proof checking (as in Dedukti).
It will be interesting to investigate if and how the advantages of these systems can be combined, e.g., if \mmt can delegate to ELPI for proof search.